# Population Redistribution among Multiple Electronic States of Molecular Nitrogen Ions in Strong Laser Fields

Jinping Yao[1], Shicheng Jiang[2], Wei Chu[1], Bin Zeng[1], Chengyin Wu[3,4,§], Ruifeng Lu[2,5,†], Ziting Li[1,6], Hongqiang Xie[1,7], Guihua Li[1], Chao Yu[2], Zhanshan Wang[6], Hongbing Jiang[3], Qihuang Gong[3,4] and Ya Cheng[1,*]

[1]*State Key Laboratory of High Field Laser Physics, Shanghai Institute of Optics and Fine Mechanics, Chinese Academy of Sciences, Shanghai 201800, China*

[2]*Department of Applied Physics, Nanjing University of Science and Technology, Nanjing 210094, China*

[3]*State Key Laboratory for Mesoscopic Physics, Department of Physics, Peking University, Beijing 100871, China*

[4]*Collaborative Innovation Center of Quantum Matter, Beijing 100871, China*

[5]*State Key Laboratory of Molecular Reaction Dynamics, Dalian Institute of Chemical Physics, Chinese Academy of Sciences, Dalian 116023, China*

[6]*School of Physics Science and Engineering, Tongji University, Shanghai 200092, China*

[7]*University of Chinese Academy of Sciences, Beijing 100049, China*

[§]*cywu@pku.edu.cn*

[†]*rflu@njust.edu.cn*

[*]*ya.cheng@siom.ac.cn*

**Abstract**

We carry out a combined theoretical and experimental investigation on the population distributions in the ground and excited states of tunnel ionized nitrogen molecules at various driver wavelengths in the near- and mid-infrared range. Our results reveal that efficient couplings (i.e., population exchanges) between the ground $N_2^+(X^2\Sigma_g^+)$ state and the excited $N_2^+(A^2\Pi_u)$ and $N_2^+(B^2\Sigma_u^+)$ states occur in strong laser fields. The couplings result in the population inversion between $N_2^+(X^2\Sigma_g^+)$ and $N_2^+(B^2\Sigma_u^+)$ states at the wavelengths near 800 nm, which is verified by our experiment by observing the amplification of a seed at ~391 nm. The result provides insight into the mechanism of free-space nitrogen ion lasers generated in remote air with strong femtosecond laser pulses.

Tunnel ionization is one of the most fundamental processes for molecules in intense laser fields when the Keldysh parameter is less than unity [1]. It naturally launches an attosecond electron bunch with a kinetic energy up to ~100 eV level, enabling probing molecular dynamics with sub-femtosecond, sub-Ångström spatio-temporal resolution based on either electron recollision induced dissociation [2-4] or laser-induced electron diffraction [5,6]. In addition, tunnel ionization rate is also highly sensitive to the ionization potentials and orbital geometries of molecules, providing new opportunities for imaging molecular orbitals [7,8]. Recently, it has been further reported that strong laser fields can simultaneously induce tunnel ionizations from several molecular orbitals because of the multi-electron effect and the close ionization potentials of the outermost and a few lower-lying orbitals [9-15]. A direct consequence of the multi-orbital ionization is that the molecular ions can be populated in not only the ground but also the excited electronic states. As a commonly accepted assumption, the molecular ions in the excited electronic states are generated through the tunnel ionization from the lower-lying orbitals, with their populations being determined by the corresponding ionization probabilities.

Since molecular ions in the excited states can decay to the ground state by spontaneous fluorescence emission, fluorescence spectroscopy has been employed to identify the electronic states of the tunnel ionized molecules and determine their population distributions, which offers an access to image the lower-lying molecular orbitals [13,15]. Owing to their higher ionization potentials, the exponential suppression of the ionization probabilities from the lower-lying orbitals suggests that the populations in the excited states should be much less than that in the ground state. Thus, it comes as a major surprise to observe a population inversion between $N_2^+(B^2\Sigma_u^+)$ and $N_2^+(X^2\Sigma_g^+)$ states when nitrogen molecules are subject to strong femtosecond laser pulses at 800-nm wavelength [16]. The implication of the novel phenomenon on the remote sensing application has been widely recognized because it provides an ultrafast, intense free-space laser source in remote air [16-21]. However, the mechanism responsible for the population inversion is still far from being understood. One may argue that the population inversion might be realized by other ionization/excitation channels, such as electron recollision induced excitation [2-4], non-adiabatic multi-electron effect [22], and electron correlation during ionization [23]. Indeed, these channels can promote the population of the excited molecular ions, but their contributions are insufficient to produce the population inversion observed for nitrogen molecules in strong femtosecond laser pulses at 800-nm wavelength. In this Letter, we present a combined theoretical and experimental investigation on the population dynamics of the tunnel ionized nitrogen molecules at the ground and excited states. Our results reveal that efficient couplings between the ground $N_2^+(X^2\Sigma_g^+)$ and the excited $N_2^+(A^2\Pi_u)$ and $N_2^+(B^2\Sigma_u^+)$ states enable generation of the population inversion between $N_2^+(X^2\Sigma_g^+)$ and $N_2^+(B^2\Sigma_u^+)$ states at pump wavelengths near 800 nm, which explains the experimental observations of the amplification of a seed at ~391 nm and clarify the mechanism of free-space nitrogen ion lasers generated in remote air with strong femtosecond laser pulses.

The experiments were carried out using a Ti:sapphire laser (Legend Elite-Duo, Coherent Inc.) and an optical parametric amplifier (OPA, HE-TOPAS, Light Conversion Ltd.). The femtosecond laser pulses (1 kHz, 800 nm, ~40 fs) from the Ti:sapphire laser were firstly split into two beams. One beam with the pulse energy of ~6.5 mJ was injected into the OPA to generate wavelength-tunable mid-infrared laser pulses in the spectral range from 1200 nm to 2400 nm, enabling performing the experiment at either 800 nm or mid-infrared pump wavelengths. The pulse energies at all wavelengths were chosen to be ~800 μJ. The other beam with a pulse energy of 300 μJ was frequency-doubled by a β-BaB$_2$O$_4$ (BBO) crystal and then served as an external seed. To obtain the strongest amplification at ~391-nm wavelength with the seed pulses, we carefully tuned the BBO crystal to slightly adjust the central wavelength of the seed pulse. The time delay between the pump and the seed pulses was controlled by a motorized linear translation stage, which was placed in the optical path of the seed beam. The seed and pump pulses were collinearly combined using a dichroic mirror and then were focused by an $f$=15cm plano-convex lens into the gas chamber filled with high-purity nitrogen gas. To minimize the seed produced by nonlinear propagation of the pump pulses (e.g., white-light or harmonic generations), we chose the relatively low gas pressure (i.e., 70 mbar) in all measurements. In particular, the external seed was polarized perpendicularly to the polarization of the pump laser at all the wavelengths, thus the self-generated emission by the pump laser can be easily removed using a Glan-Taylor prism placed in front of the spectrometer. The output signal was then focused into a slit and completely captured by a grating spectrometer (Shamrock 303i, Andor) whereas the residual pump laser was removed using a stack of band-pass filters.

Figure 1(a) shows the forward spectrum of the external seed injected into the plasma channel generated by the 800-nm pump pulse with an optimized delay (blue dashed line). For comparison, the spectrum of initial seed pulse is also presented by a red solid line. Clearly, in the presence of the 800-nm pump pulse, a narrow-bandwidth peak with a central wavelength at ~391 nm appears on top of the spectrum of the seed pulse, which corresponds to the transition of $N_2^+(B^2\Sigma_u^+, v'=0) \rightarrow N_2^+(X^2\Sigma_g^+, v=0)$. The strong narrow-bandwidth 391-nm emission is ascribed to a seed-amplified $N_2^+$ laser owing to the population inversion between $N_2^+(B^2\Sigma_u^+, v'=0)$ and $N_2^+(X^2\Sigma_g^+, v=0)$ states. We stress here that the two spectra in Fig. 1(a) are recorded by collecting all the photons in the signal beams into the spectrometer, providing unambiguous evidence on the gain of the external seed during its propagation in the plasma channel. The observation clearly suggests that at the 800-nm pump wavelength, population inversion has occurred in the tunnel ionized nitrogen molecules.

Next, we performed the same measurement using the wavelength-tunable mid-infrared pump pulses from the OPA. At multiple pump wavelengths ranging from 1200 nm to 2000 nm, the experimental results are similar as shown by the typical spectra recorded at pump wavelengths of 1750 nm and 1960 nm in Fig. 1(b) and (c). At the pump wavelength of 1750 nm, neither the third nor the fifth harmonics covers the frequency of the transition of $N_2^+(B^2\Sigma_u^+, v'=0) \to N_2^+(X^2\Sigma_g^+, v=0)$. Whereas the fifth harmonic of 1960 nm pump laser covers the transition wavelength of ~391 nm, which offers a self-generated seed source. To focus on the influence of the pump wavelength, we managed to maintain the same experimental parameters (e.g., pump energy, focusing geometry, gas pressure, seed wavelength, seed energy, etc.) as those in Fig. 1(a). Interestingly, different from the result obtained with 800-nm pump pulses, the spectra of the external seed pulses show a significant absorption peak at ~391 nm when the pump wavelengths are set at 1750 nm and 1960 nm, as illustrated in Fig. 1(b) and (c), respectively. This observation indicates that more ions are populated at the ground $N_2^+(X^2\Sigma_g^+, v=0)$ state as compared to the excited $N_2^+(B^2\Sigma_u^+, v'=0)$ state, and thus no population inversion is built up at the two pump wavelengths. The comparison shows that the population inversion responsible for the 391-nm $N_2^+$ laser has a strong dependence on the pump wavelength.

It is noteworthy that previously, the narrow-bandwidth laser-like emission at ~391 nm has been observed without the injection of the external seed at the pump wavelengths in Fig. 1(a-c). The origin of the emission obtained with the mid-infrared pump sources is yet to be identified, which will be a subject of the future investigation. Below, we will focus our attention on the mechanism behind the population inversion generated at 800-nm pump wavelength. Specifically, we have measured the gain dynamics of the 391-nm emission as a function of the time delay between the pump and the seed pulses, as shown in Fig. 1(d). It was found that the gain increases rapidly on a time scale of ~300 fs followed by a relatively slow decay on the time scale of ~1 ps, which is consistent with our previous measurement [24].

To gain deeper insight into the underlying mechanism of the experimental observations, we perform time-dependent quantum wave packet calculations to simulate the dynamic processes of nitrogen molecular ions in strong laser fields. Here, we assume the ionization occurs near the peak of the pump laser pulses. Tunnel ionizations from different molecular orbitals generate molecular nitrogen ions in different electronic states, whose vibrational populations are determined by Franck-Condon transition from neutral nitrogen molecules to the molecular nitrogen ions in corresponding electronic states [25]. Meanwhile, it is known that tunnel ionized molecules may populate in the excited states due to the photoelectron recollision induced excitation [26]. According to a time dependent wave-packet calculation, the excitation probability of molecular ions can reach ~20% with the recollision photoelectrons [4]. Thus we assume that 20% of molecular nitrogen ions populate in the $A^2\Pi_u$ state and the $B^2\Sigma_u^+$ state each. After the generation of nitrogen molecular ions, the remaining laser field will cause the coupling of the electronic states and promote the population exchanges among them. The time-dependent Schrödinger equation for describing the couplings among these electronic states is given by:

$$i\frac{\partial}{\partial t}\begin{pmatrix}\Psi_X(R,t)\\ \Psi_A(R,t)\\ \Psi_B(R,t)\end{pmatrix}=[-\frac{1}{2\mu}\frac{\partial^2}{\partial R^2}+V(R,t)]\begin{pmatrix}\Psi_X(R,t)\\ \Psi_A(R,t)\\ \Psi_B(R,t)\end{pmatrix}, \qquad (1)$$

where atomic units are used, $\mu$ is the reduced mass and $R$ is the internuclear separation of $N_2^+$. The potential matrix $V(R, t)$ for the three-state system can be written as:

$$V(R,t)=\begin{pmatrix} V_X(R) & \vec{\mu}_{XA}(R)\cdot\vec{E}(t) & \vec{\mu}_{XB}(R)\cdot\vec{E}(t) \\ \vec{\mu}_{XA}(R)\cdot\vec{E}(t) & V_A(R) & 0 \\ \vec{\mu}_{XB}(R)\cdot\vec{E}(t) & 0 & V_B(R) \end{pmatrix}, \qquad (2)$$

where the diagonal elements respectively denote the Born-Oppenheimer potential energy curves (PECs) of $N_2^+(X^2\Sigma_g^+)$, $N_2^+(A^2\Pi_u)$ and $N_2^+(B^2\Sigma_u^+)$. The off-diagonal elements denote the coupling terms induced by laser fields under dipole approximation with $\vec{\mu}_{XA}(R)$ and $\vec{\mu}_{XB}(R)$ being the electronic transition moments (TMs) between the relevant pairs of electronic states. $\vec{E}(t)$ represents the linearly polarized laser field with a constant amplitude for the first three optical cycles and a Gaussian falling edge of 12 cycles. The laser frequency and peak intensity (i.e., $\sim 2\times 10^{14}$ W/cm$^2$) are taken as the same as the experimental parameters. Because the photoionization dominantly occurs near the peak of the pump laser pulse, it is reasonable to choose this laser profile when the strong field-induced population transfer is calculated.

Using the MOLPRO package [27], we perform state-of-the-art *ab initio* calculations for PECs and TMs. Specifically, Dunning's aug-cc-pVQZ basis set, active space including 13 valence orbitals which correlate with the 2s, 2p, 3s, 3p orbitals of N atom, and Davidson correction for higher-order correlation were employed in the calculations of the state-averaged complete active space self-consistent-field/multireference configuration interaction (SA-CASSCF/MRCI). Figure 2(a) shows the PECs of $N_2^+(X^2\Sigma_g^+)$, $N_2^+(A^2\Pi_u)$, $N_2^+(B^2\Sigma_u^+)$ and $N_2(X^1\Sigma_g^+)$. It should be noted that $\vec{\mu}_{XA}(R)$ is perpendicular to the molecular axis and $\vec{\mu}_{XB}(R)$ parallel to the molecular axis. There is no dipole moment between $N_2^+(A^2\Pi_u)$ and $N_2^+(B^2\Sigma_u^+)$ states. In the simulation, we included all the possible orientations of molecular axis and averaged them with angular dependent ion yields. As shown in Fig. 2(b), efficient population exchanges are observed between the ground and the two excited electronic states at 800-nm pump wavelength. The population mainly distributes in the ground electronic state of $N_2^+$ immediately after the ionization, even the photoelectron recollision induced excitation is taken into account. The remaining laser fields cause population transfer between the ground and the excited electronic states. At the end of the laser pulse, the population of $N_2^+(B^2\Sigma_u^+)$ is higher than that of $N_2^+(X^2\Sigma_g^+)$. We further explore the vibrational distribution of each electronic state of $N_2^+$ by projecting the final wave packets onto the field-free vibrational eigenfunctions. Figure 2(c) shows the vibrational distribution of $N_2^+(B^2\Sigma_u^+)$. It can be seen that the excited ions are mainly populated in the ground vibrational state. The results demonstrate that the population inversion has been formed between $N_2^+(B^2\Sigma_u^+, v'=0)$ and $N_2^+(X^2\Sigma_g^+, v=0)$ states. The population inversion leads to the amplification of the seed pulse at ~391 nm, which agrees with our experimental observation in strong laser field of 800-nm wavelength. Last but not least, it should be noted that although there is no direct coupling between $N_2^+(A^2\Pi_u)$ and $N_2^+(B^2\Sigma_u^+)$ states, the participation of $N_2^+(A^2\Pi_u)$ state is indispensable for generating the population inversion between excited $N_2^+(B^2\Sigma_u^+)$ state and ground $N_2^+(X^2\Sigma_g^+)$ state. As evidenced by Fig. 2(d), if we remove the coupling between $N_2^+(A^2\Pi_u)$ and $N_2^+(X^2\Sigma_g^+)$ states, the molecular ions will be dominantly populated in $N_2^+(X^2\Sigma_g^+)$ but not $N_2^+(B^2\Sigma_u^+)$ state. By comparing Fig. 2(b) with Fig. 2(d), we conclude that the involvement of $N_2^+(A^2\Pi_u)$ not only depletes the population of $N_2^+(X^2\Sigma_g^+)$, but also increases the population of $N_2^+(B^2\Sigma_u^+)$. The combined effects lead to the population inversion between $N_2^+(B^2\Sigma_u^+)$ and $N_2^+(X^2\Sigma_g^+)$.

In our simulations, we found that the generation of population inversion relies on the initial population distribution of molecular nitrogen ions. Without taking the photoelectron recollision induced excitation into account, the population inversion between $N_2^+(B^2\Sigma_u^+)$ and $N_2^+(X^2\Sigma_g^+)$ cannot be achieved purely through the population transfer induced by the laser field. Moreover, the generation of population inversion also relies on the pump wavelength. As shown in Fig. 3(a) and 3(b), the couplings between the ground state and excited electronic states are very weak when the pump wavelengths are 1750 nm and 1960 nm and population inversion remains unachieved. Generally speaking, at the peak intensity and pulse duration in Fig. 2(b), population inversion cannot be achieved with the pump wavelengths in the range of 1200 – 2000 nm. This also agrees with our experimental results.

Based on our experimental observation and theoretical calculation, we provide the following physical picture on the generation of population inversion in the tunnel ionized nitrogen molecules in strong laser fields. As illustrated in Fig. 4(a), most nitrogen molecules remain neutral in the rising edge of the pump laser field, because tunnel ionization is highly sensitive to the field strength. When approaching the peak of envelop of the driver laser, the molecules are rapidly ionized and populated to various electronic states. According to the laser parameters of our experiment, the molecular nitrogen ions are mainly distributed in the ground vibrational state of $N_2^+(X^2\Sigma_g^+)$. After the generation of the molecular nitrogen ions, as shown in Fig. 4(b), population transfers occur between the ground and the two excited electronic states due to the couplings induced by the strong laser field [28]. Population inversion between $N_2^+(B^2\Sigma_u^+)$ and $N_2^+(X^2\Sigma_g^+)$ is finally achieved with the assistance of $N_2^+(A^2\Pi_u)$ as discussed above. Furthermore, our theoretical simulations show that the population inversion can be achieved for a broad range of the laser parameters (data not shown), which is consistent with the large number of experimental observations on the strong-field induced $N_2^+$ laser reported so far [16,17,20,21].

To conclude, we have performed experimental and theoretical investigations on the populations of tunnel ionized nitrogen molecules with strong laser fields. It is generally known that tunnel ionization from lower-lying orbitals can lead to direct generation of molecular nitrogen ions in the excited states, but their populations are much less than that in the ground state. Inclusion of the photoelectron recollision can greatly promote the excitation efficiency, whereas the populations in the excited states are still too low to accomplish the population inversion required for initiating any lasing actions. However, due to the couplings of the ground and excited states induced by the strong laser fields, efficient population exchanges between the ground and excited electronic states occur, which play a crucial role in the generation of either spontaneous fluorescence or laser emissions from the molecular nitrogen ions. Our finding not only unlocks the puzzle of the strong field induced air lasing in tunnel ionized nitrogen molecules, but also has important implication for investigating the excited state dynamics of molecular ions in intense laser fields.

The authors would like to thank Dr. Aijie Zhang for her assistance on the ab initio calculation. This work is supported by the National Basic Research Program of China (Grant Nos. 2013CB922403 and 2014CB921303), and National Natural Science Foundation of China (Grant Nos. 11134010, 60921004, 21373113, 11474009, 11434002, 11204332, 11304330 and 11404357).

Note added - *In a recent complementary work, Liu et al. experimentally demonstrated the recollision excitation plays a significant role for initiating lasing actions in tunnel ionized nitrogen molecules* [29].

**Captions of figures:**

Fig. 1 (Color online) (a) The forward spectra obtained by injecting an external seed pulse into the plasma channel induced by (a) 800-nm, (b) 1750-nm and (c) 1960-nm pump pulses (blue dashed lines). For comparison, the spectra of the seed pulses are indicated by red solid lines in Fig. (a-c). (d) The temporal evolution of the 391-nm coherent emission with the time delay between the pump and seed pulses at the pump wavelength of 800 nm.

Fig. 2 (Color online) (a) Potential energy curves of the electronic states obtained by cubic spline interpolation of ab initio data using the MOLPRO package. (b) The temporal evolution of the population distributions in the electronic states of $N_2^+$ at the pump wavelength of 800 nm. (c) The temporal evolution of vibrational states distribution of $N_2^+(B^2\Sigma_u^+)$ in the 800-nm laser field. (d) The temporal evolution of the population distributions in $N_2^+(X^2\Sigma_g^+)$ and $N_2^+(B^2\Sigma_u^+)$ states obtained by removing $N_2^+(A^2\Pi_u)$ states.

Fig. 3 (Color online) The temporal evolution of the population distributions calculated at the pump wavelengths of (a) 1750 nm and (b) 1960 nm.

Fig. 4 (Color online) Schematic diagram of the pumping mechanism for establishing the population inversion. In region I of the pump pulse, photoionization is weak and most molecules remain neutral. (a) Near the peak of envelop of the driver laser, the molecules are populated to various electronic states of $N_2^+$ by tunnel ionization (region II, in orange color). (b) After tunnel ionization, populations are redistributed among the three electronic states of $N_2^+$ through efficient couplings between the ground and the excited states (region III, hatched area of the pump pulse).

**Fig. 1**

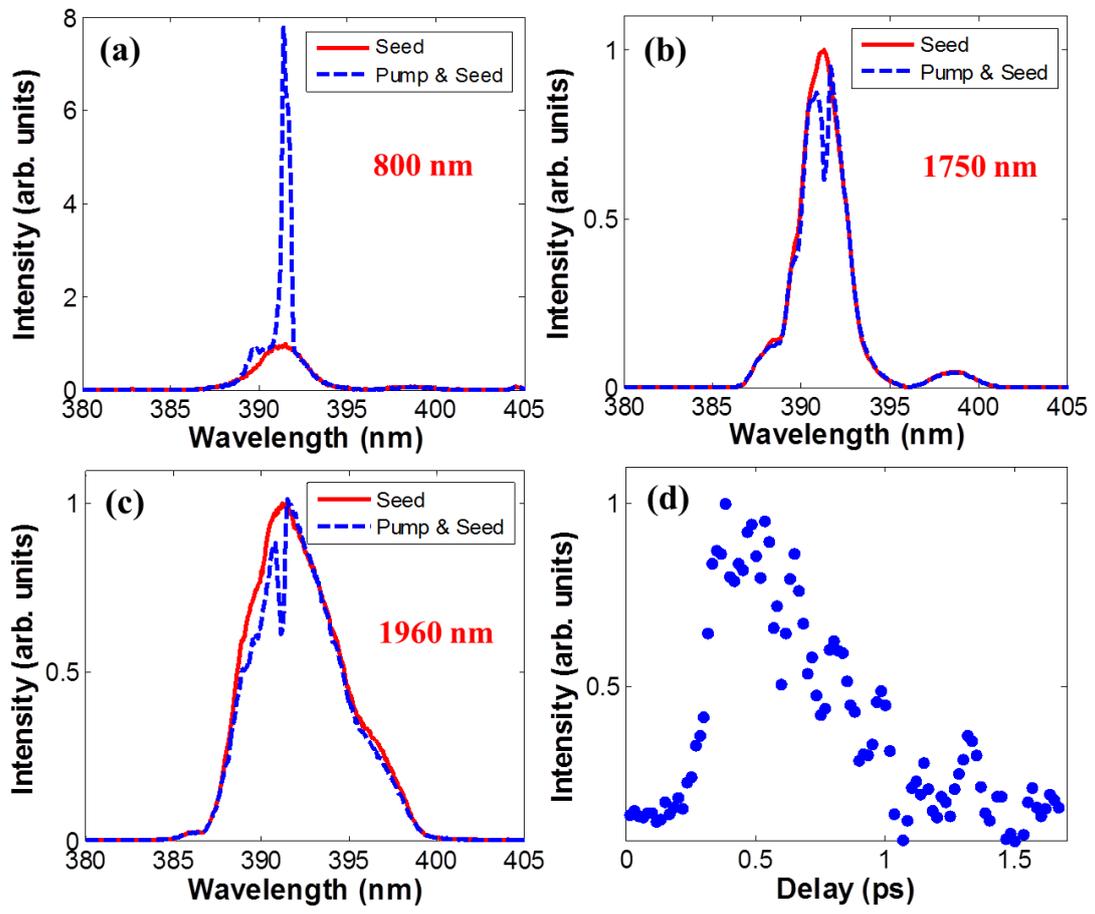

**Fig. 2**

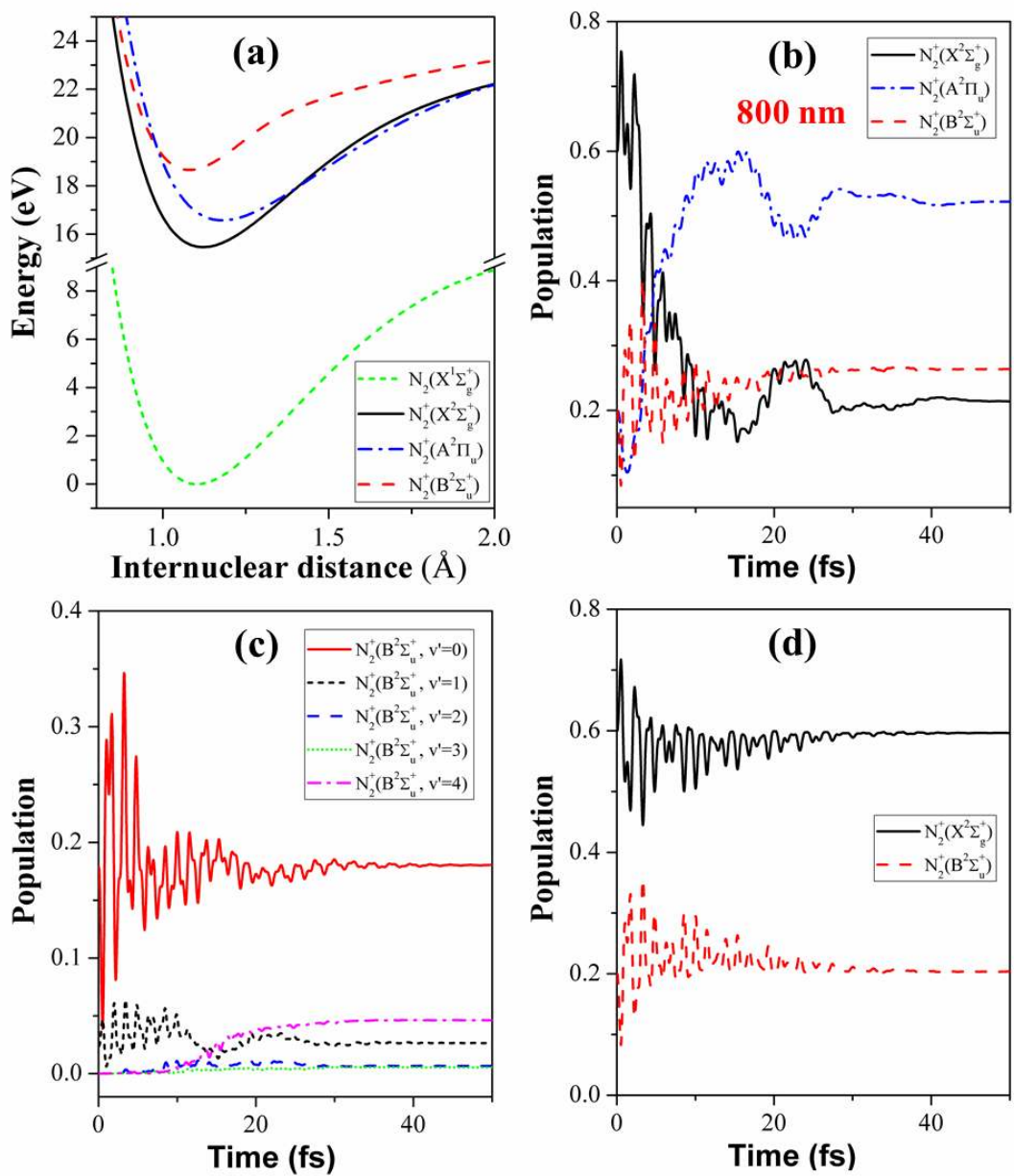

**Fig. 3**

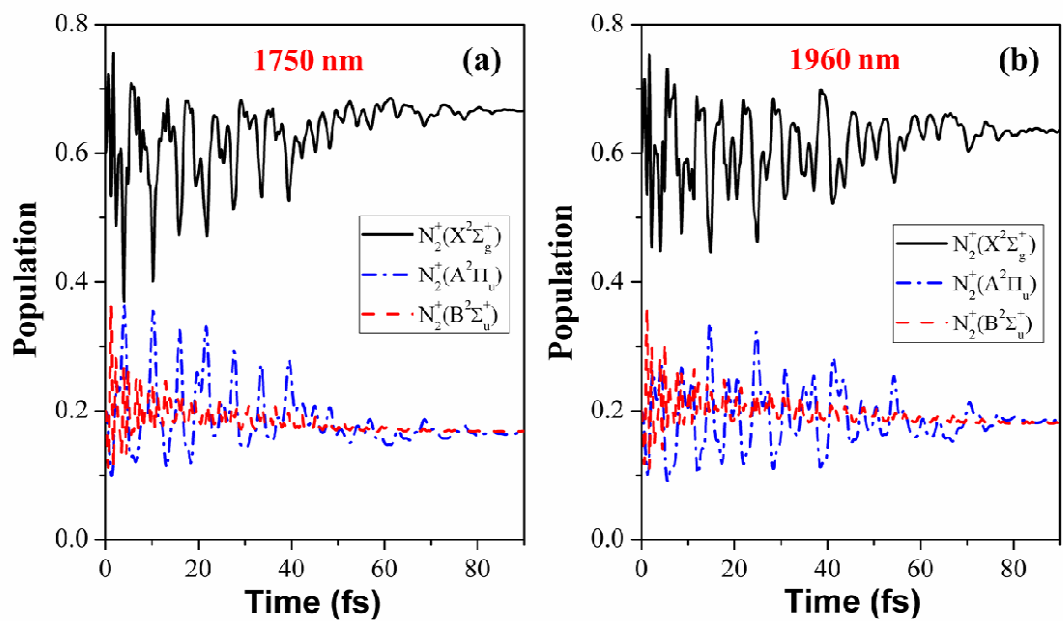

**Fig. 4**

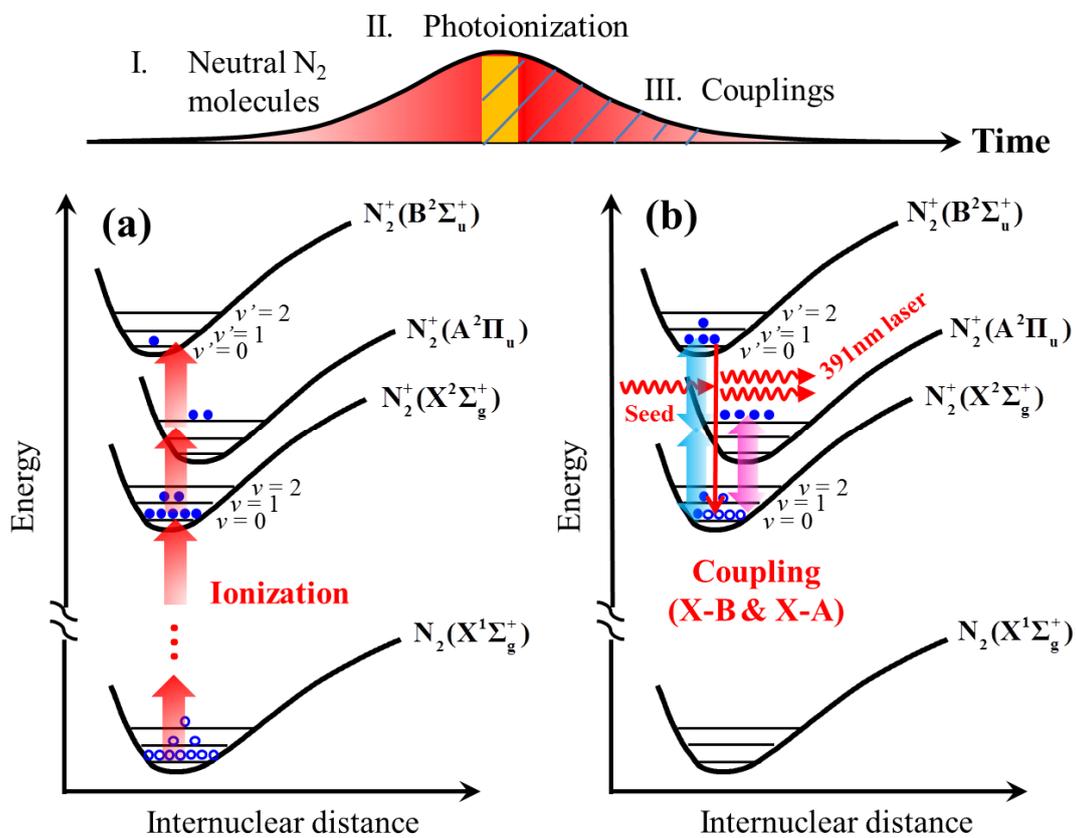